\documentclass[aps,pra,superscriptaddress,showpacs,twocolumn]{revtex4}
\usepackage{units}
\usepackage{amsmath}
\usepackage{amssymb}
\usepackage{graphicx}
\usepackage{bm}
\usepackage{multirow,color,relsize,ulem}

\newcommand{\be}{\begin{equation}}
\newcommand{\ee}{\end{equation}}
\newcommand{\e}{\varepsilon}
\newcommand{\bj}[2]{J_{#1}(#2)}
\newcommand{\bh}[2]{H_{#1}(#2)}

\newcommand{\bjs}[1]{J_{#1}}
\newcommand{\bhs}[1]{H_{#1}}
\newcommand{\bjps}[1]{J'_{#1}}
\newcommand{\bhps}[1]{H'_{#1}}

\newcommand{\cs}[1]{\cos(#1\theta)}
\newcommand{\re}{\text{Re}}
\newcommand{\im}{\text{Im}}
\newcommand{\ang}{\text{Arg}}
\newcommand{\largetriangleleft}{\mathlarger{\mathlarger{\mathlarger{\triangleleft}}}}
\newcommand{\largetriangleright}{\mathlarger{\mathlarger{\mathlarger{\triangleright}}}}
\newcommand{\wf}{\psi}
\newcommand{\wfb}{\varphi}

\begin{document}
\title{Multimode Coupling by Boundary Wave Scattering}

\author{Li Ge}
\email{lge@princeton.edu}
\affiliation{Department of Electrical Engineering, Princeton University, Princeton, NJ 08544, USA}
\author{Qinghai Song}
\affiliation{Department of Electronic and Information Engineering, Shenzhen Graduate School, Harbin Institute of Technology, Shenzhen, 518055, China}
\author{Brandon Redding}
\affiliation{Department of Applied Physics, Yale University, New Haven, CT 06520-8482, USA}
\author{Alexander Ebersp\"acher}
\affiliation{Institut f{\"u}r Theoretische Physik, Universit{\"a}t Magdeburg, Postfach 4120, D-39016 Magdeburg, Germany}
\author{Jan Wiersig}
\affiliation{Institut f{\"u}r Theoretische Physik, Universit{\"a}t Magdeburg, Postfach 4120, D-39016 Magdeburg, Germany}
\author{Hui Cao}
\email{hui.cao@yale.edu}
\affiliation{Department of Applied Physics, Yale University, New Haven, CT 06520-8482, USA}
\date{\today}

\begin{abstract}
We show that coupling among multiple resonances can be conveniently introduced and controlled by boundary wave scattering. We demonstrate this principle in optical microcavities of quasi-circular shape, where the couplings of multiple modes are determined by the scattering from different harmonic boundary deformations. We analyze these couplings using a perturbation theory, which gives an intuitive understanding of the first-order and higher-order scattering processes.
Different scattering paths between two boundary waves can either enhance or reduce their coupling strength.
The effect of controlled multimode coupling is most pronounced in the direction of output from an open cavity, which can cause a dramatic change of the external cavity field distribution.
\end{abstract}
\pacs{42.25.Gy, 42.55.Sa, 03.65.Nk}

\maketitle

\section{Introduction}
\label{sec:intro}
Eigenmodes are fundamental in understanding all quantum and wave phenomena. Their couplings occur when the orthogonality or biorthogonality of the system is modified,
which can be introduced, for example, by matter-mediated interaction in cavity quantum electrodynamics (cQED) \cite{cQED}, 
by nonlinearity in multimode lasers \cite{Science08}, and by linear scattering from a local defect \cite{Wiersig_PRA06} or a gradual boundary deformation in microcavities \cite{Wiersig_PRL06}. While predictive models based on the Jaynes-Cummings Hamiltonian \cite{JC} and Maxwell-Bloch equations \cite{MB} can be employed to capture the effect of mode coupling in cQED systems and multimode lasers, scattering-induced coupling between two eigenmodes is usually described by phenomenological models \cite{Heiss}. Progresses have been made recently in understanding mode coupling due to the scattering from a minute boundary deformation \cite{Bogomolny,Wiersig_PRA12,e3}, and quasi-degeneracy was shown to be an important factor that can lead to a drastic change of the outcoupling direction in an open system \cite{e3}.

One important aspect that has not been addressed systematically is the mutual coupling of multiple modes due to linear scattering. In this report we show that such {\it multimode} coupling can be conveniently achieved and controlled by boundary wave scattering.
This approach applies to a general eigenvalue problem
\be
\left[-\nabla^2+V(\vec{r})\right]\wf(\vec{r})=E\wf(\vec{r}),\label{eq:H}
\ee
which can be realized, for example, in a vibrating membrane \cite{Perez}, a dielectric microcavity \cite{Chang, Vahala}, an optical trap for exciton-polariton condensate \cite{Polariton}, and a quantum dot \cite{Reed}. The scalar eigenmodes $\wf(\vec{r})$ represent the vibrational amplitude, components of the electromagnetic wave, or the probability wave function in the corresponding physical systems.

Below we exemplify the properties of boundary wave scattering in an open quasi-circular cavity, with $V(\vec{r})=-(n^2-1)E$ inside and $V(\vec{r})=0$ outside. Eq.~(\ref{eq:H}) then becomes the scalar two-dimensional Helmholtz equation, which describes, for example,
the propagation of transverse electric (TE) or transverse magnetic (TM) waves in a dielectric cavity of refractive index $n$.
For simplicity, we assume the cavity shape is nearly circular and symmetric along the horizontal axis $\theta=0,180^\circ$, and we describe the cavity boundary using harmonic series
\be
\rho(\theta)\equiv R\left[1+ \sum_{\nu}\e_\nu \cs{\nu}\right]\quad(\nu=2,3,\ldots), \label{eq:rho}
\ee
where $R$ is the average radius of the cavity. The dipolar term ($\e_1 \cos\theta$) is not included because it mostly leads to a lateral shift of the cavity if $|\e_1|\ll1$ and can be eliminated by choosing a proper origin.
The harmonic boundary deformations in (\ref{eq:rho}) can be employed as individual turning knobs to introduce and control coupling among multiple modes of different angular momenta, which is a generalization of the procedure introduced in Ref.~\cite{e3}.
Such a scheme can be utilized to alter the outcoupling direction of an eigenmode deterministically, through first-order and higher-order boundary wave scattering.

This report is organized as follows. In Sec.~\ref{sec:pert} we review the perturbation theory for the scalar Helmholtz equation in a quasi-circular cavity. We relate each perturbation contribution in the presence of the harmonic deformations in (\ref{eq:rho}) to scattering strengths of different orders.
In Sec.~\ref{sec:single_pert} we demonstrate multimode coupling via a {\it single} harmonic boundary deformation using TM modes.
In Sec.~\ref{sec:multi_pert} we examine how {\it multiple} harmonic boundary deformations can be introduced to control couplings among TM modes, and we show that the same procedure also applies for TE modes. The summary is given in Sec.~\ref{sec:conclusion}, in which we also comment on the similarity of boundary wave scattering to other quantum and wave phenomena involving multipath interference.

\section{Perturbation thoery}
\label{sec:pert}

We begin by reviewing the perturbation theory for TM \cite{Bogomolny} and TE modes \cite{e3} of the scalar Helmholtz equation in a quasi-circular system. In the absence of deformation, each eigenmode $\wfb$ of the circular system is characterized by its angular momentum $m$ and radial quantum number $\eta$. The latter indicates the number of intensity peaks in the radial direction inside the cavity, and we will refer to the modes with $\eta \ll m$ as the boundary waves, since they are confined closely to the inside of the cavity boundary.
For convenience, we represent their angular dependence by sine and cosine functions, i.e.
\be
\wfb_{m,\eta}(\vec{r}) \propto
\begin{cases}
J_m(nK_{m,\eta}r)\cs{m},\\
J_m(nK_{m,\eta}r)\sin(m\theta),
\end{cases}
\ee
inside the cavity. Outside the cavity $\wfb_{m,\eta}(\vec{r})$ are similarly defined, with the Bessel functions $J_m(nK_{m,\eta}r)$ replaced by the Hankel functions of the first kind $H_m(K_{m,\eta}r)$ and properly normalized to guarantee the continuity of $\wfb_{m,\eta}(\vec{r})$ at the cavity boundary. The complex resonant frequencies $K_{m,\eta}$, corresponding to the square root of $E$ in Eq.~(\ref{eq:H}), are determined by the resonance condition
\be
S_m(Z) \equiv n\frac{\bjps{m}(nZ)}{\bjs{m}(nZ)}-\frac{\bhps{m}(Z)}{\bhs{m}(Z)} = 0
\ee
for TM modes and
\be
T_m(Z) \equiv \frac{1}{n}\frac{\bjps{m}(nZ)}{\bjs{m}(nZ)}-\frac{\bhps{m}(Z)}{\bhs{m}(Z)} = 0
\ee
for TE modes, where $Z\equiv KR$.

The eigenmodes inside the cavity are slightly perturbed in the presence of a minute boundary deformation.
The perturbed modes each have a dominant angular momentum $m$ and recognizable radial quantum number $\eta$, and they are still parity eigenstates about the horizontal symmetry axis if the boundary takes the form of Eq.~(\ref{eq:rho}). We denote them $\wf_{m,\eta}$ and their resonant frequencies $k_{m,\eta}$ to distinguish them from the unperturbed modes $\wfb_{m,\eta}$ and their frequencies $K_{m,\eta}$. Below we focus on the even-parity modes, and the analysis for the odd-parity modes is similar. We keep the indices of $\wf$, $\wfb$, $k$ and $K$ only when they are important. Using the ansatz
\be
\wf(\vec{r})=\begin{cases}
\sum_{p\geq0} a_p \frac{\bj{p}{nkr}}{\bj{p}{nkR}}\cs{p}, & r<\rho(\theta), \\
\sum_{p\geq0} (a_p+b_p)\frac{\bh{p}{kr}}{\bh{p}{kR}}\cs{p}, & r>\rho(\theta),
\end{cases}\label{eq:ansatz}
\ee
and $a_m\equiv1$ for the dominant angular momentum, Dubertrand {\it et al.} found that in a cavity with the boundary given by
\be
\rho(\theta) = R[1+\lambda f(\theta)],\, f(\theta)=f(-\theta),
\ee
the perturbed quantities are
\begin{align}
kR &= Z\bigg[ 1 - \lambda A_{mm} - \lambda^2\bigg(Z(n^2-1)\sum_{q\neq m}A_{mq}A_{qm}\frac{1}{S_q} \nonumber \\
&\quad - \frac{3A_{mm}^2-B_{mm}}{2}  - Z (A_{mm}^2-B_{mm})\frac{H'_m}{H_m} \bigg)\bigg], \label{eq:k0}\\
a_p &= \lambda Z(n^2-1)\frac{1}{S_p}\bigg( A_{pm} +  \lambda\bigg\{A_{pm}A_{mm}\left(\frac{Z}{S_p}S_p'-1\right) \nonumber \\
&\quad + \frac{B_{pm}}{2}\left[1+Z\left(\frac{H'_m}{H_m} + \frac{H'_p}{H_p}\right)\right] \nonumber \\
&\quad + Z(n^2-1)\sum_{q\neq m}A_{pq}A_{qm}\frac{1}{S_q}\bigg\}\bigg), \label{eq:ap1}\\
b_p &=  \frac{\lambda^2Z^2}{2}(n^2-1)B_{pm} \label{eq:bp}
\end{align}
up to order $O(\lambda^2)$ for TM modes \cite{Bogomolny}. We have dropped the argument $Z$ in $S$, $H$, and their derivatives $S'$, $H'$ with respect to $Z$. The coefficients $A_{pm},B_{pm}$ are given by
\begin{align}
A_{pm} &\equiv \frac{c_p}{\pi} \int_0^{\pi} f(\theta)\cs{p}\cs{m} d\theta, \\
B_{pm} &\equiv \frac{c_p}{\pi} \int_0^{\pi} f^2(\theta)\cs{p}\cs{m} d\theta,
\end{align}
with $c_{p}=2~(p>0), 1~(p=0)$.

The perturbation theory was extended to TE modes in Ref.~\cite{e3}, which is more complicated due to the discontinuity of the radial derivative of $\wf$ at the cavity boundary. The results up to the order $O(\lambda)$ are given by
\begin{align}
kR &= Z\left[ 1 - \lambda A_{mm}\right], \\
a_p &= \lambda Z\left[S_m\left(\frac{\bhps{p}}{\bhs{p}}-\frac{\bhps{m}}{\bhs{m}}\right)-T'_m\right]\frac{1}{T_p}A_{pm},\label{eq:ap_TE}\\
b_p &= \lambda ZS_m A_{pm},
\end{align}
and we note that $b_p$ now has a first-order term in $\lambda$ and $kR$ is the same as for TM modes to this order.

The creation of angular momentum sidebands $a_{p\neq m}$ inside the cavity can be considered as the result of boundary wave scattering.
The perturbation results above give an intuitive understanding of these scattering processes. Take the TM polarization as example, $a_p$ given by Eq.~(\ref{eq:ap1}) can be rearranged as
\be
a_p = \alpha_{pm} + \sum_{q\neq m} \alpha_{pq}\alpha_{qm} + (\dots)B_{pm} + (\dots)A_{pm}A_{mm},\label{eq:ap2}
\ee
where
\be
\alpha_{pm} = \lambda Z(n^2-1)\frac{1}{S_p} A_{pm}\label{eq:alpha}
\ee
can be considered as the scattering strength for the first-order process $m\rightarrow p$ by the $\cos(m\pm p)\theta$ deformation in (\ref{eq:rho}). In our notation the angular momenta $m,p$ of the boundary waves are non-negative, representing the clockwise (CW) wave with a positive sign and counterclockwise (CCW) wave with a negative sign. The $\cos(m-p)\theta$ deformation scatters the CW (CCW) wave of angular momentum $m$ ($-m$) into the CW (CCW) wave of angular momentum $p$ ($-p$), and the $\cos(m+p)\theta$ deformation scatters the CW (CCW) wave of angular momentum $m$ ($-m$) into the CCW (CW) wave of angular momentum $p$ ($-p$). $\alpha_{pm}$ is proportional to $S_{p}^{-1}(K_{m,\eta}R)$, to which we will refer the spectral function. If there is another resonance $K_{p,\eta'}$ in the vicinity of $K_{m,\eta}$, we then find $S_{p}(K_{m,\eta}R)\approx S_{p}(K_{p,\eta'}R) = 0$ and the spectral function can become very large, leading to a dramatic sensitivity of the scattering strength to the deformation \cite{e3}. In the appendix we show that for high-$Q$ modes of $\eta=1$, this sensitivity from a low-order harmonic boundary deformation maximizes in the mesoscopic regime that lies in the crossover between the wavelength regime ($\lambda\sim R$) and the semiclassical limit ($\lambda\ll R$)], where $\lambda$ is the wavelength.

The second term in Eq.~(\ref{eq:ap2}) represents different paths that consist of two successive first-order scattering processes, i.e. $m\rightarrow q \rightarrow p$ for all $q\neq m$. Such second-order processes depend not only on the spectral function of the final state [i.e. $S_{p}^{-1}(K_{m,\eta}R)$] but also on the spectral function of the intermediate state [i.e. $S_{q}^{-1}(K_{m,\eta}R)$]. There is another type of second-order processes in $a_p$, which is represented by the term proportional to $B_{pm}$ in Eqs.~(\ref{eq:ap1}),(\ref{eq:ap2}). Their scattering strengths do not depend on the spectral function of the intermediate state, thus we will refer to them as ``virtual'' processes. The last $O(\lambda^2)$ term in Eq.~(\ref{eq:ap2}) has a more complicated dependence on the spectral function of the final state. Its scattering strength is proportional to $A_{mm}$, which represents the scattering of the CW and CCW waves of angular momentum $\pm m$ into each other by the $\cs{2m}$ deformation.

\section{Multimode coupling via a single harmonic boundary deformation}
\label{sec:single_pert}

Henceforth we exemplify multimode coupling using the TM polarization unless specified otherwise. Let us first consider a single harmonic perturbation $\cs{\nu}$ with an amplitude $|\e_\nu|\ll1$. As we have discussed in the previous section, this boundary deformation scatters the boundary wave of angular momentum $m$ into two sidebands $m\pm \nu$ to the leading order, whose amplitudes are given by
\be
\alpha_{m\pm \nu,m} = \frac{\e_\nu(n^2-1)Z}{2S_{m\pm \nu}} \label{eq:alpha0}
\ee
from Eq.~(\ref{eq:alpha}).
This boundary wave scattering introduces coupling between $\wf_{m,\eta}$ and two other modes $\wf_{m+\nu,\eta'},\wf_{m-\nu,\eta''}$, whose dominant angular momentum is $m+\nu$ and $m-\nu$, respectively.

\begin{figure}[t]
\centering
\includegraphics[width=\linewidth]{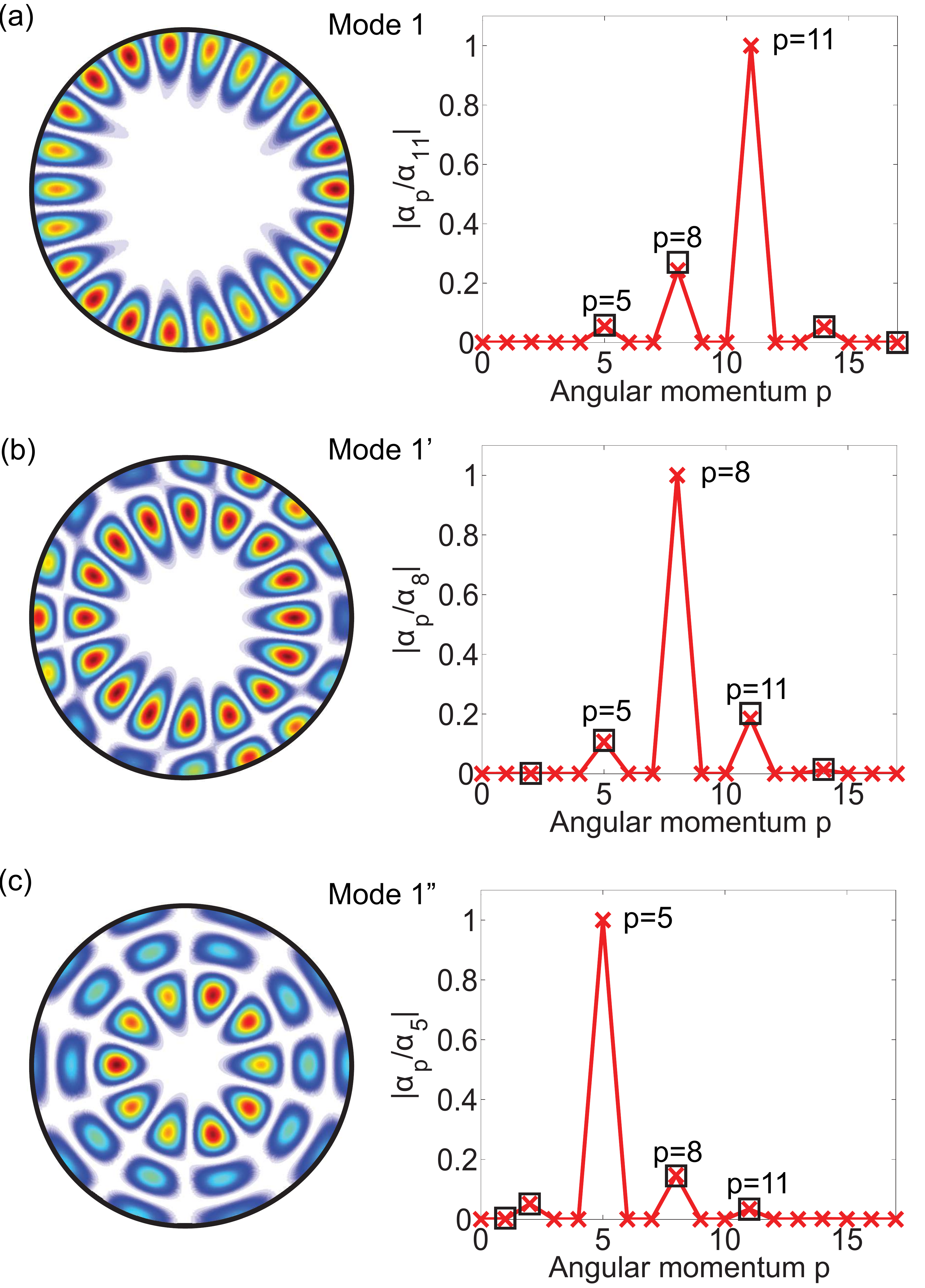}
\caption{(Color online) (a) Intracavity intensity distribution and angular momentum components $|a_p|$ of mode 1  in a quasi-circular cavity $\rho(\theta)\equiv R[1+ 0.01 \cs{3}]$. Its resonant frequency is $k_{11,1}R = 4.593 - 1.603\times10^{-4}i$. Connected crosses show the numerical values of $|\alpha_p|$ and black squares show the perturbation results from Eqs.~(\ref{eq:alpha0}),(\ref{eq:beta}). (b,c) Same as (a) for two nearby modes of frequency $k_{8,2}R=4.699-1.351\times10^{-3}i$ and $k_{5,3}R=4.515 - 4.046\times10^{-2}i$.} \label{fig:e3}
\end{figure}

Higher-order scattering processes create weaker sidebands at $m\pm2\nu,m\pm3\nu,\ldots$, coupling more modes with decreasing strength in general \cite{bibnotes1}. The strength $\beta_{m\pm2\nu,m}$ of the second-order scattering can be found in Eq.~(\ref{eq:ap2}):
\begin{align}
\beta_{m\pm2\nu,m} &\approx \alpha_{m\pm 2\nu, m\pm \nu}\alpha_{m\pm \nu, m}. \label{eq:beta}
\end{align}
We have assumed $\nu<m$, with which $A_{mm}$ and the last term in Eq.~(\ref{eq:ap2}) vanish. We have also neglected the ``virtual'' process given by $B_{pm}$ in Eq.~(\ref{eq:ap2}); it is weak compared with the right hand side of Eq.~(\ref{eq:beta}), since it does not depend on the spectral function $S_{m\pm\nu}^{-1}$ of the intermediate state, which needs to be large for the second-order scattering strength to be non-negligible.

In Fig.~\ref{fig:e3} we show one example with a $\cs{3}$ boundary deformation in a circular cavity of index $n=3.13$. Near $Z = 4.6$ there are three eigenmodes of angular momenta $m=11,m'=8,m''=5$ and radial quantum number $\eta=1, \eta'=2, \eta''=3$.
They are calculated using a scattering-matrix method similar to that described in Refs. \cite{Narimanov, Tureci}.
We refer to them as modes 1, $1'$, and $1''$, and they have increasing cavity decay rates, defined by $\kappa=-2\im[Z]>0$. We first focus on mode 1 and point out that the two first-order sidebands of mode 1 at $p=8,14$ do not have equal strength; the presence of mode $1'$ leads to a spectral function $S_8^{-1}>S_{14}^{-1}$, and the sideband at $p=8$ is about four times stronger than that at $p=14$. Similarly, the proximity of mode $1''$ enhances the scattering into the second-order sideband of mode 1 at $p=5$, which is even stronger than the first-order one at $p=14$.

\begin{figure}[t]
\centering
\includegraphics[width=\linewidth]{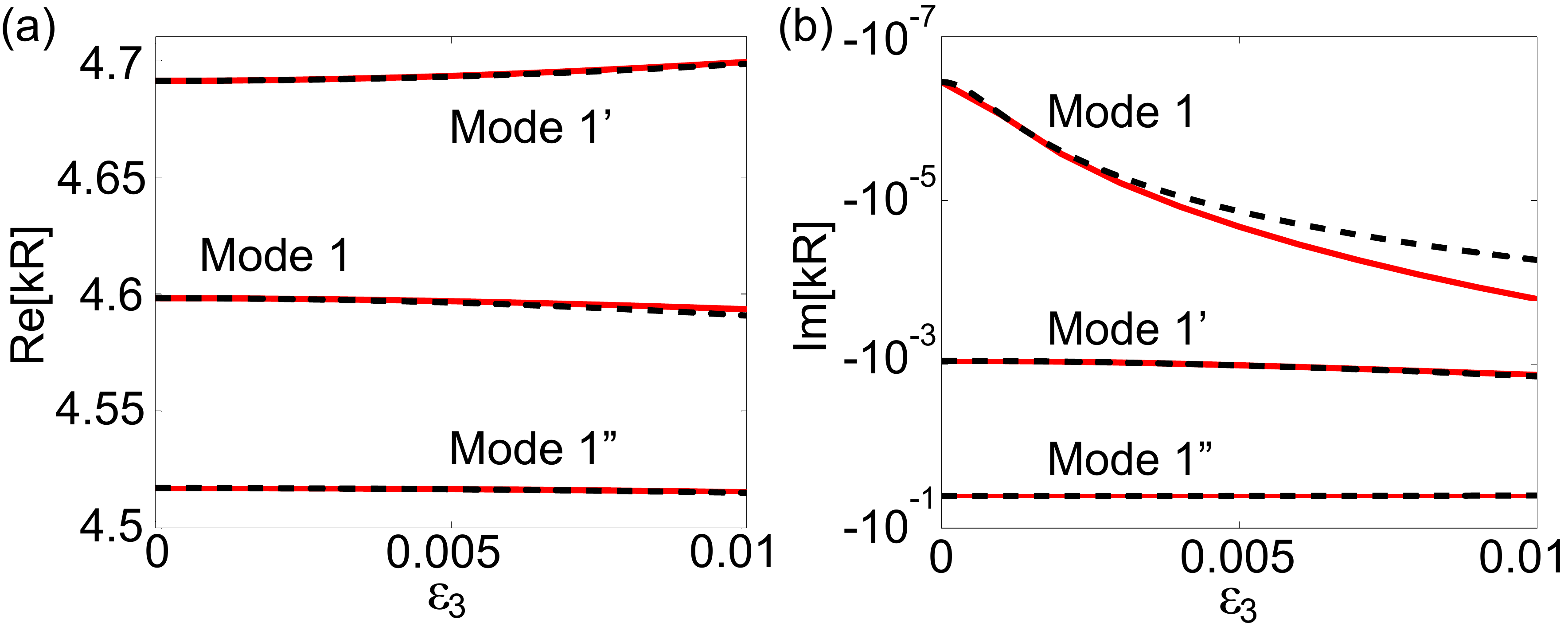}
\caption{(Color online) Real (a) and imaginary (b) parts of the complex frequencies of modes 1, $1'$, and $1''$ in Fig.~\ref{fig:e3}. Solid lines and black dashed lines show the numerical data and the second-order perturbation result from Eq.~(\ref{eq:k}), respectively.}\label{fig:e3_vas}
\end{figure}

Figs.~\ref{fig:e3}(b,c) for modes $1'$ and $1''$ further display the mutual coupling between them and mode 1. The couplings, however, affect these eigenmodes differently. For example, their $\im[kR]$ all vary on the scale of $10^{-4}$ when $\e_3$ changes from 0 to 0.01, captured by a quadratic function of $\e_3$
\be
kR =  Z\left(1-\frac{\alpha_{m-3,m} + \alpha_{m+3,m}}{2}\e_3\right),\label{eq:k}
\ee
where $\alpha_{m-3,m},\alpha_{m+3,m}$ are given by Eq.~(\ref{eq:alpha0}) and linear in $\e_3$. Eq.~(\ref{eq:k}) is derived from Eq.~(\ref{eq:k0}) using the fact that both $A_{mm}$ and $B_{mm}$ vanish in this example.
Such a variation, however, barely changes the cavity decay rates of mode $1'$ and $1''$ but increases that of mode 1 by more than two orders of magnitude (see Fig.~\ref{fig:e3_vas}).

Another example is their intracavity field distribution. Close to the cavity boundary the wave function (\ref{eq:ansatz}) inside the cavity can be approximated by
\be
\wf_m(R,\theta)\approx\cs{m} + \sum_{p=m\pm \nu}\alpha_{pm}\cs{p}, \label{eq:intraInt}
\ee
with the higher-order sidebands neglected. The phases of $\alpha_{pm}$ determine the field distribution and its orientation. For example, we find that both $\alpha_{8,11},\alpha_{14,11}$ are almost real and {\it positive} at a positive $\e_3$ from Eq.~(\ref{eq:alpha0}). We then expect both sidebands at $p=8,14$ in mode 1 to interfere constructively with the dominant angular component $m=11$ at $\theta\approx0^\circ,120^\circ,240^\circ$, resulting in a ``$\largetriangleright$" shape. Similarly, Eq.~(\ref{eq:alpha0}) predicts that $\alpha_{5,8},\alpha_{11,8}$ are almost real but {\it negative} at a positive $\e_3$. The beating of the $p=11,5$ sidebands in mode $1'$ with the dominant $m'=8$ component then leads to an enhanced field intensity at $\theta\approx60^\circ,180^\circ,300^\circ$, leading to a ``$\largetriangleleft$" orientation. These predictions are confirmed by the numerical calculations shown in Fig.~\ref{fig:e3}.

\section{Controlling Multimode coupling via multiple harmonic boundary deformations}
\label{sec:multi_pert}

\begin{figure}[b]
\centering
\includegraphics[width=\linewidth]{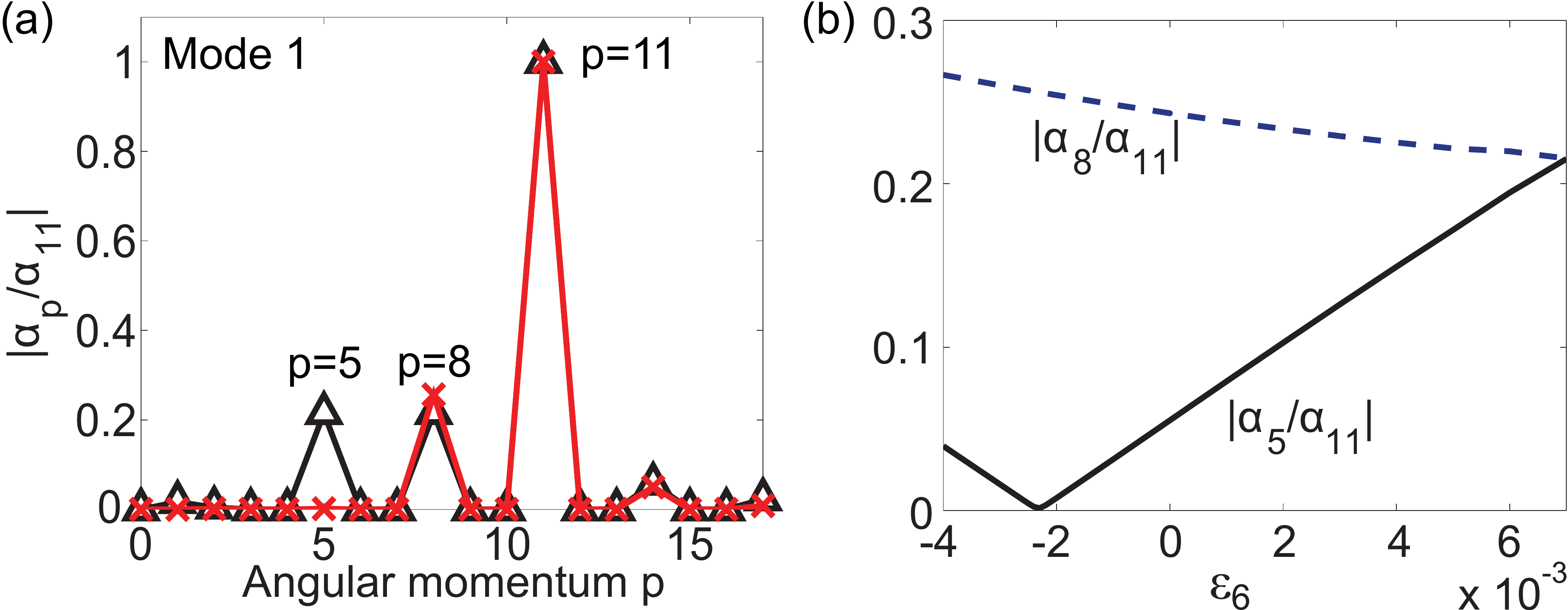}
\caption{(Color online) Modification of mode 1 in Fig.~\ref{fig:e3} by adding an $\e_6 \cs{6}$ deformation to the cavity boundary. (a) Angular momentum components $|\alpha_p|$ of mode 1 inside the cavity at $\e_6=-0.0023$ (connected red crosses) and 0.006 (connected black triangles). (b) $|\alpha_5|$ and $|\alpha_8|$ of mode 1 as a function of $\e_6$. $\e_3$ is fixed at 0.01. } \label{fig:e6}
\end{figure}

The above example illustrates how the boundary wave scattering from a {\it single} harmonic boundary deformation couples multiple eigenmodes. The couplings of mode $1'$ to modes 1, $1''$ are of first-order and are stronger than that between modes 1 and $1''$, which is of second-order. To vary the latter without affecting the former, one can introduce extra couplings by adding additional harmonic boundary deformations. For example, a $\cs{6}$ deformation adds first-order coupling between modes 1, $1''$, which can be tuned to reduce or enhance their existing coupling due to the $\cs{3}$ deformation. As shown in Fig.~\ref{fig:e6}, the $p=5$ sideband in mode 1 is almost canceled completely at $\e_6=-0.0023$ and raised to about the same height with the $p=8$ sideband at $\e_6=0.007$. All other angular components only have a minute change with $\e_6$ due to higher-order perturbations.

The tunability of the total coupling between two modes depends on the phases of the individual couplings from different scattering paths. To find  the general requirement that a first-order scattering $m\rightarrow p$ cancels a second-order scattering $m\rightarrow q \rightarrow p$, we again turn to Eq.~(\ref{eq:ap2}) and employ the same approximation used in deriving Eq.~(\ref{eq:beta}), which leads to
\be
a_p \approx \alpha_{pm} + \alpha_{pq}\alpha_{qm}. \label{eq:ap3}
\ee
From Eqs.~(\ref{eq:ap1}) and (\ref{eq:alpha0}) we know that $\alpha_{pm}$ and $\alpha_{pq}$ above are proportional to the same spectral function of the final state, i.e. $S^{-1}_p$. $\alpha_{pm}$ and $\alpha_{pq}$ are then in-phase ($\pi$-out-of-phase) if $\e_{m-p}$ and $\e_{q-p}$ have the same sign (opposite signs). Therefore, the requirement for the aforementioned cancelation at some value of $\e_{m-p}$ is to have a real $\alpha_{qm}$.
Indeed we find $\ang[\alpha_{8, 11}]=0.008$ in the example given above, where $m=11$, $q=8$, and $p=5$, and a negative $\e_6=-0.0023$ is needed to cancel the $p=5$ sideband at a positive $\e_3=0.01$. We also note that the phase of $a_p$ changes by about $\pi$ across $\e_6=-0.0023$ as a result.

The effect of controlled multimode coupling is most pronounced in the outcoupling of the high-$Q$ modes. Recent studies \cite{Wiersig_PRA06, Song_PRL10} show that the outcoupling direction of a high-$Q$ mode can be completely overwhelmed by that of a lower-$Q$ mode to which it couples. The situation becomes more interesting if the high-$Q$ mode couples to more than one lower-$Q$ mode, such as the case in Fig.~\ref{fig:e3}.
Unlike the intracavity intensity distribution which is largely determined by the dominant angular momentum and the strong first-order sidebands [see Eq.~(\ref{eq:intraInt}) and its discussion], the weaker sidebands of lower angular momenta can also have a strong influence on the outcoupling due to their stronger leakiness. More specifically, the wave function (\ref{eq:ansatz}) in the farfield becomes
\begin{align}
\wf(r\rightarrow\infty) &\propto \sum_p (a_p+b_p)\frac{e^{-ip\pi/2}}{H_p(kR)}\cs{p} \nonumber \\
&\equiv\sum_pW_p\cs{p}, \label{eq:farfield}
\end{align}
using the large-argument asymptotic form of the Hankel function of the first kind
\be
H_p(z\rightarrow\infty)\rightarrow\sqrt{\frac{2}{\pi z}}e^{i(z-p\pi/2-\pi/4)}.
\ee
We note that the amplitude of $H_p(kR)$ in the denominator of Eq.~(\ref{eq:farfield}), evaluated at the average radius of the cavity instead of the farfield, reduces dramatically for a smaller angular momentum $p$, which represents the stronger leakiness mentioned above.

\begin{figure}[t]
\centering
\includegraphics[width=0.85\linewidth]{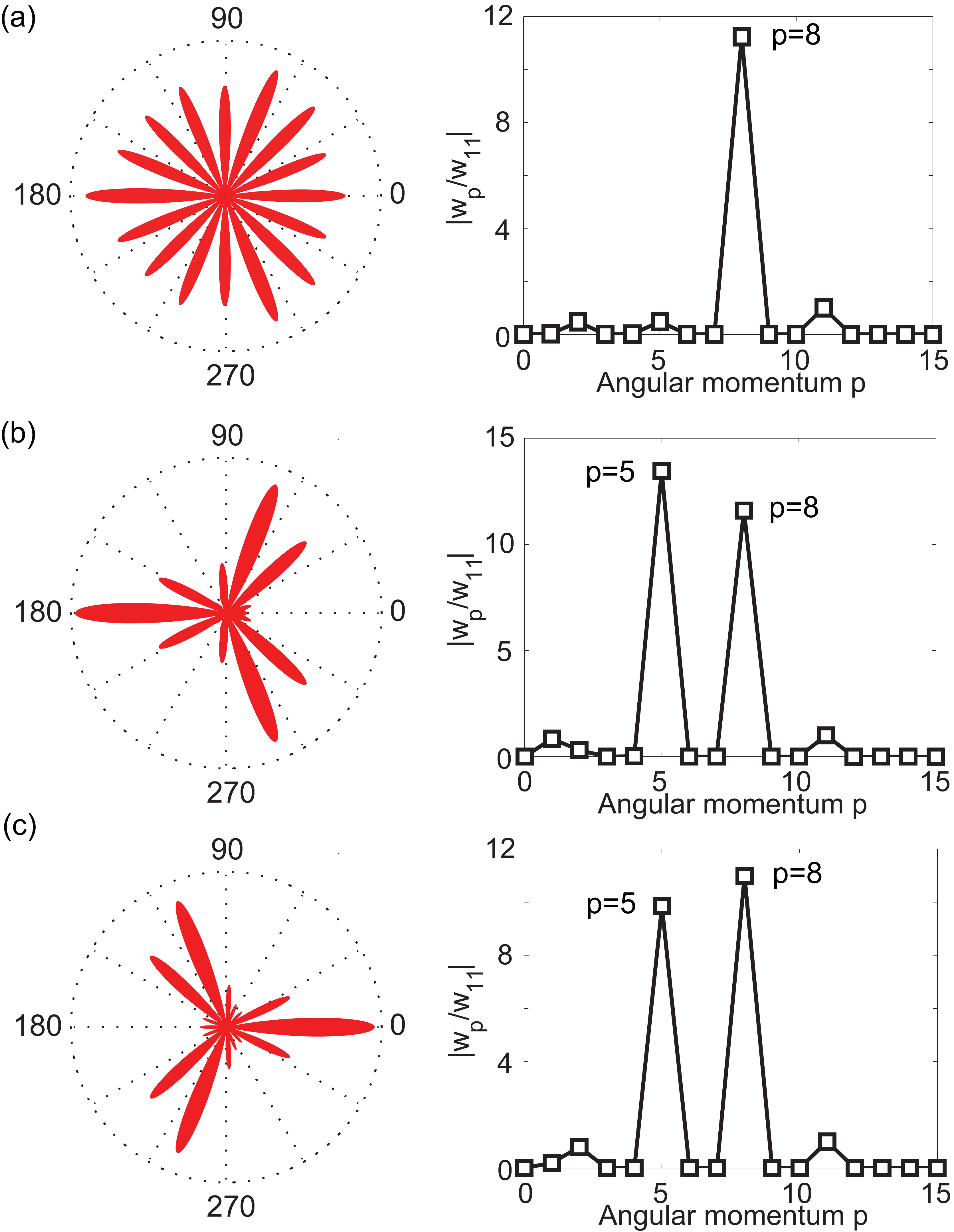}
\caption{(Color online) (a,b,c) Farfield intensity pattern (left) and its angular components (right) of mode 1 in Fig.~\ref{fig:e3}(a) at $\e_6=-0.0023$, -0.004, and -0.001. $\e_3$ is fixed at 0.01. } \label{fig:e6_outside}
\end{figure}

\begin{figure}[b]
\centering
\includegraphics[width=\linewidth]{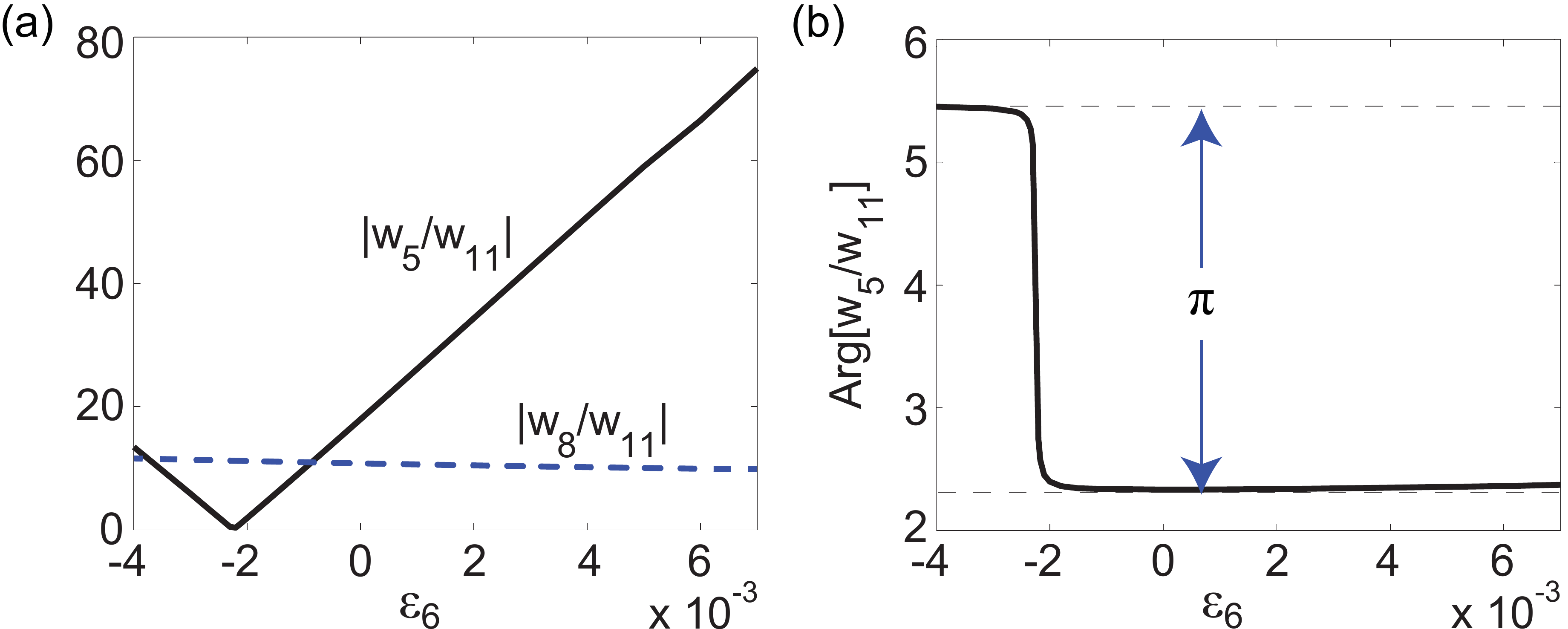}
\caption{(Color online) (a) Angular components $|W_5|$ and $|W_8|$ of mode 1 outside the cavity as a function of $\e_6$. (b) $\pi$-phase jump of $W_5$ near $\e_6=-0.0023$. The two dashed lines are separated by $\pi$ and used as references. $\e_3$ is fixed at 0.01. } \label{fig:B_e6_outside}
\end{figure}

Now let us examine how the outcoupling direction of mode 1 changes with $\e_6$ in the example shown Fig.~\ref{fig:e6}. At $\e_6=-0.0023$ the $m=5$ sideband outside the cavity is very small, similar to what happens inside the cavity. The outcoupling of mode 1 is then dominated by the $p=8$ sideband, which leads to an approximate angular dependence of $\cs{16}$ [Fig.~\ref{fig:e6_outside}(a)]. As $\e_6$ changes from -0.0023, the cancelation of the two scattering paths is removed and the coupling between modes 1 and $1''$ increases rapidly; the $p=5$ sideband outside the cavity becomes comparable to the $p=8$ sideband at $\e_6\approx-0.004,-0.001$ [see Fig.~\ref{fig:B_e6_outside}(a)], which are about ten times as large as the $m=11$ component, the dominant one inside the cavity.
Depending on the relative phase and amplitude of $W_5$ and $W_8$, the beating of these two largest angular components can lead to a quite different outcoupling direction. For example, using Eqs.~(\ref{eq:ap1}),(\ref{eq:bp}) we find that $W_5,W_8$ are approximately $\pi$-out-of-phase at $\e_6\approx-0.004$, and the outcoupling is enhanced in the $\theta\approx60^\circ,180^\circ,300^\circ$ directions [Fig.~\ref{fig:e6_outside}(b)]. At $\e_6\approx-0.001$ however, the phase of $W_5$ changes roughly by $\pi$ [see Fig.~\ref{fig:B_e6_outside}(b)]. This is because $W_5$ is approximately proportional to $a_5$ since $|a_5|\gg|b_5|$, and we know from our discussion of Eq.~(\ref{eq:ap3}) that the phase of $a_5$ jumps by about $\pi$ across $\e_6=-0.0023$. Meanwhile, $W_8$ varies little for such a small change of the minute $\e_6$, since it depends on $\e_6$ only through a second-order scattering path $11\rightarrow5\rightarrow8$. As a result, $W_5$ and $W_8$ are now approximately in phase, and the outcoupling is enhanced in the $\theta\approx0^\circ,120^\circ,240^\circ$ directions instead [Fig.~\ref{fig:e6_outside}(c)], which is flipped vertically from that at  $\e_6\approx-0.004$. In this process the intracavity intensity distribution of mode 1 barely changes from the ``$\largetriangleright$" pattern shown in Fig.~\ref{fig:e3}(a), since the modified $p=5$ sideband inside the cavity is very weak [see Fig.~\ref{fig:e6}(b)].
Thus the flipping of the outcoupling direction with $\e_6$ is different from that reported in Ref.~\cite{e3}, which involves the flipping of the intracavity field pattern as well.
As $\e_6$ moves further away from $-0.0023$, the $p=5$ sideband in mode 1 gradually becomes the dominant angular momentum outside the cavity [Fig.~\ref{fig:B_e6_outside}(a)], and the angular dependence of the outcoupling approaches $\cs{10}$ (not shown).

\begin{figure}[t]
\centering
\includegraphics[width=0.75\linewidth]{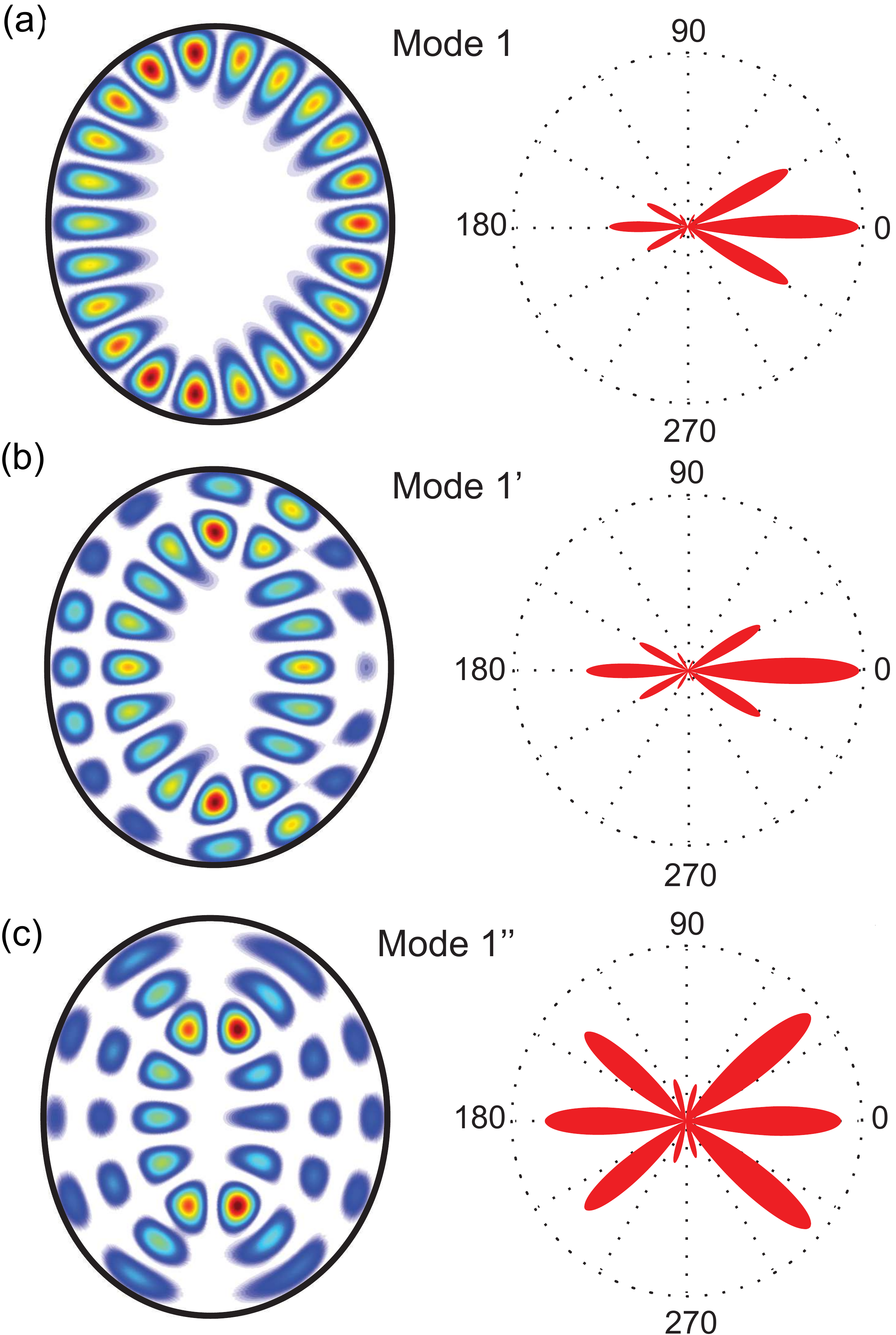}
\caption{(Color online) (a) Intracavity intensity distribution (left) and farfield intensity pattern (right) of a TE mode at $kR=4.895 - 7.272\times10^{-4}i$ in a quasi-circular cavity with $\e_2=-0.07$ and $\e_3=0.008$. (b,c) The same for two nearby TE modes at $kR=4.991 - 2.230\times10^{-2}i$ (mode $1'$) and $kR=4.866 - 0.1018i$ (mode $1''$).
} \label{fig:e6_TE}
\end{figure}

\begin{figure}[t]
\centering
\includegraphics[width=0.75\linewidth]{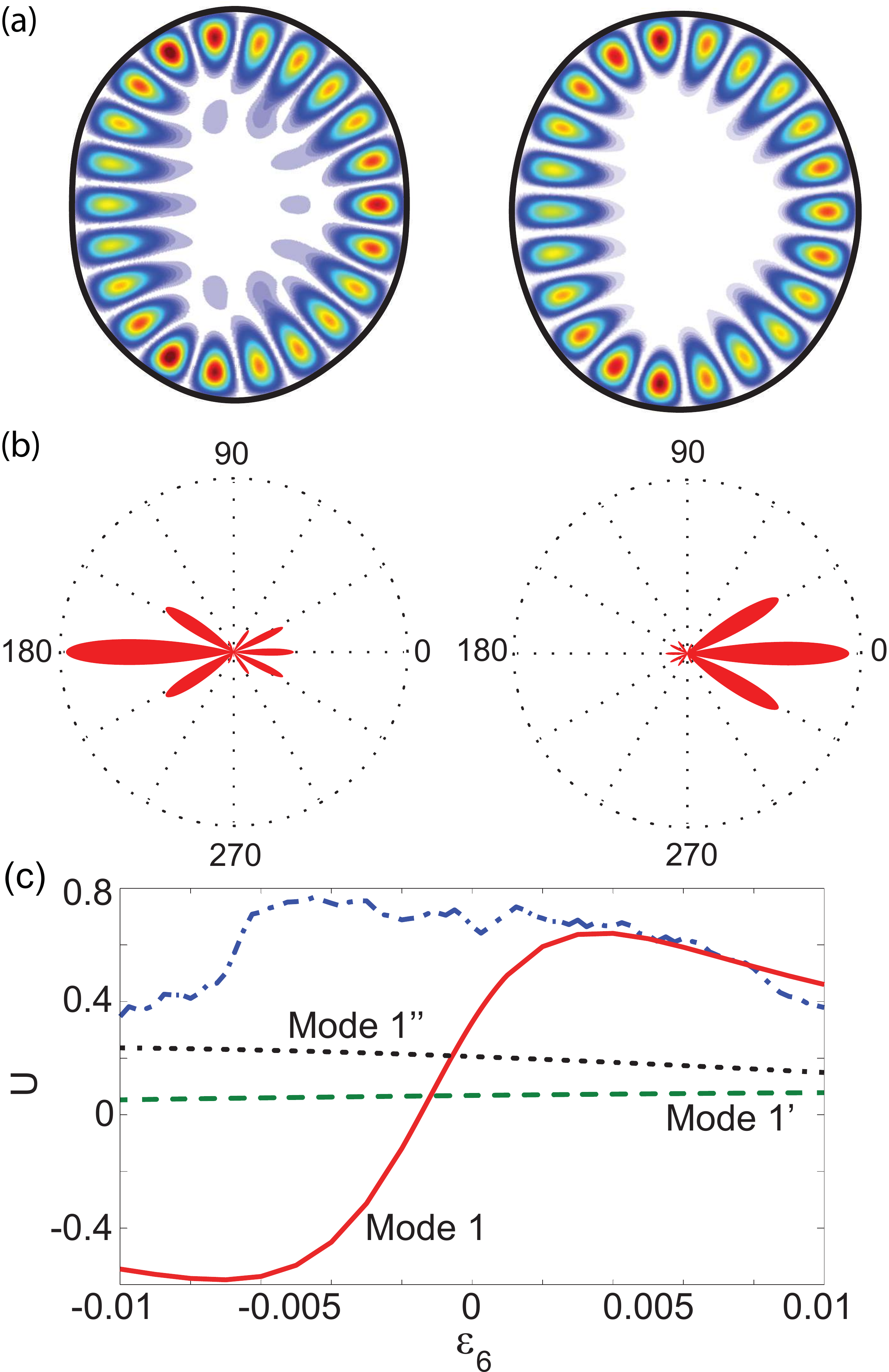}
\caption{(Color online) (a) Intracavity intensity distribution of mode 1 in Fig.~\ref{fig:e6_TE} at $\e_6=-0.008$ (left) and 0.004 (right). (b) shows their farfield intensity patterns. (c) Measure of the outcoupling direction $U$ for all three modes in Fig.~\ref{fig:e6_TE} as a function of $\e_6$.
The dash-dotted line shows the result of classical ray-tracing simulation as a comparison, which is obtained by following 50,000 initial rays uniformly distributed in the Poincar\'e surface of section \cite{rayTracing1,rayTracing2,rayTracing3} for each value of $\e_6$.
} \label{fig:U_e6_TE}
\end{figure}

To demonstrate the generality of boundary wave scattering, below we show another example using the TE polarization and with more harmonic terms in the boundary deformation. We start with $\e_2=-0.07$ and $\e_3=0.008$, and three nearby modes with $m=11,m'=8,m''=5$ can be found around $kR=4.9$ (see Fig.~\ref{fig:e6_TE}); they are the TE correspondence of the TM modes we have discussed in Figs.~\ref{fig:e3}-\ref{fig:e6_outside}, albeit the deformations are different.
We find that mode 1 couples strongly to mode $1'$, both having $W_6$ as the largest component outside the cavity; in mode $1'$ it comes from the scattering of the dominant component $\alpha_8$ inside the cavity off the large $\cs{2}$ deformation. Consequently, the outcoupling direction of mode 1 is almost identical to that of mode $1'$ [Fig.~\ref{fig:e6_TE}(a)(b)].

\begin{figure}[t]
\centering
\includegraphics[width=\linewidth]{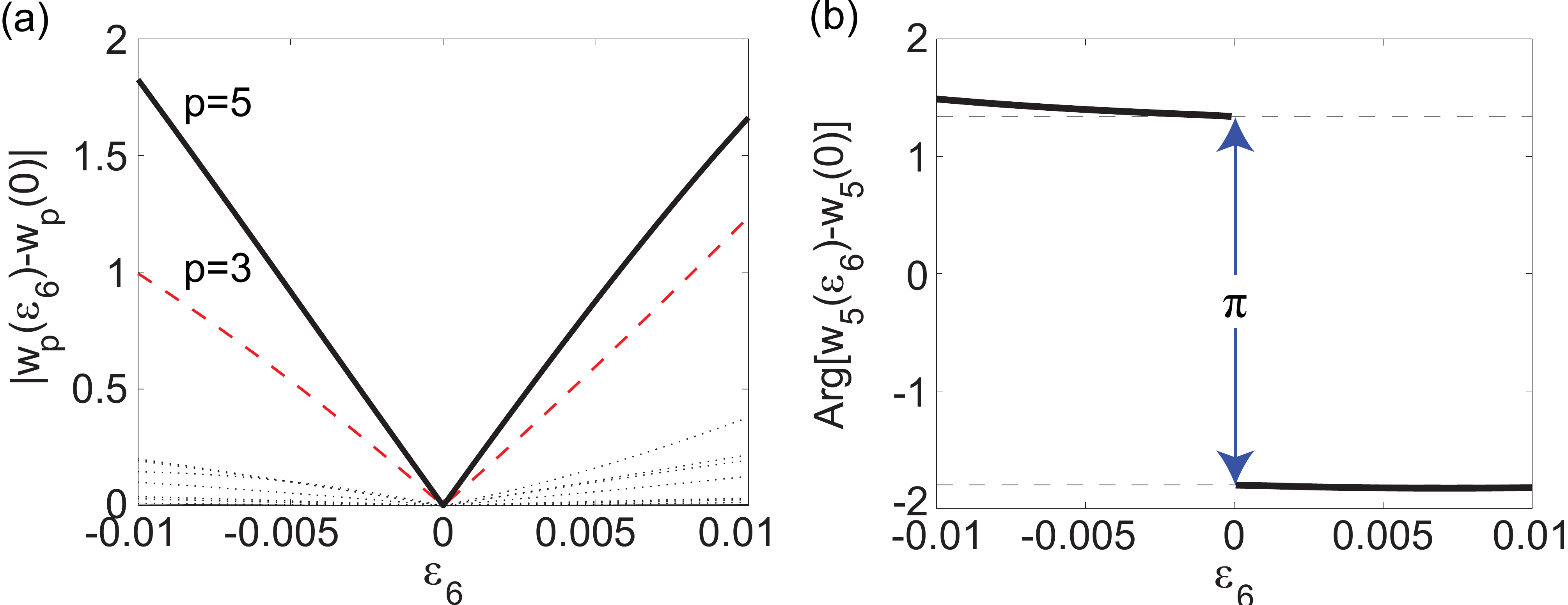}
\caption{(Color online) (a) Amplitude change to the angular components $W_p$ in mode 1 outside the cavity as a function of $\e_6$. Black solid line and red dashed line represent $W_5$ and $W_3$, and low-lying dotted lines show the rest of $W_p$ up to $p=12$. $W_6$ is normalized to be 1. $|W_5(\e_6)-W_5(0)|$ reflects the amplitude of the first-order scattering strength $11\rightarrow5$.
(b) Phase change to $W_5$ in mode 1 as a function of $\e_6$. It is fixed on both sides of $\e_6=0$ to a good approximation and jumps by $\pi$ across $\e_6=0$. The two dashed lines are separated by $\pi$ and used as references.} \label{fig:B_e6_TE}
\end{figure}

To enhance the coupling between mode 1 and $1''$, we introduce a $\cs{6}$ perturbation, similar to what is done in Figs.~\ref{fig:e6} and \ref{fig:e6_outside}. We again find that the outcoupling direction of mode 1, here indicated by
\be
U \equiv \frac{\int_0^{2\pi} I(\theta)\cos\theta d\theta}{\int_0^{2\pi}  I(\theta) d\theta}
\ee
to measure its ``skewness'' along the horizontal direction, changes dramatically from left-pointing ($U<0$) to right-pointing ($U>0$), while the outcoupling directions of modes $1'$ and $1''$ barely change [Fig.~\ref{fig:U_e6_TE}(b,c)]. We also perform a classical ray-tracing calculation of $U$ for various cavity deformations; Fig.~\ref{fig:U_e6_TE}(c) \cite{rayTracing1,rayTracing2,rayTracing3} clearly shows it does not capture the correct deformation dependence of the outcoupling direction of mode 1, which is a wave interference effect not taken into account in the classical ray dynamics.
As shown in Fig.~\ref{fig:U_e6_TE}(a), the weak field intensity of mode 1 near the cavity center at $\e_6=-0.008$ is similar to that of mode $1''$ in Fig.~\ref{fig:e6_TE}(c), which is already a hint that the aforementioned change of mode 1 is indeed caused by the newly introduced first-order coupling to mode $1''$. To further confirm this relation, we note that the amplitudes of $W_5,W_3$ in mode 1 vary most noticeably and linearly with $\e_6$ [see Fig.~\ref{fig:B_e6_TE}(a)], which are exactly the two most significant components in mode $1''$ outside the cavity. In addition, the change of $W_5$ from its value at $\e_6=0$ has almost fixed phases for $\e_6<0$ and $\e_6>0$, which differ by $\pi$ [Figs.~\ref{fig:B_e6_TE}(c)]. These observations indicate that the change of $W_5$ is of the first-order in $\e_6$, which is what we expect from the first-order scattering amplitude $a_5$ due to the $\cs{6}$ deformation, given by the TE perturbation result from Eq.~(\ref{eq:ap_TE}).

The coupling between modes 1 and $1''$ can also be enhanced via a second-order scattering process. By introducing a $\cs{4}$ perturbation and utilize the large $|\e_2|$ deformation, the scattering from $m=11$ to $p=5$ is efficiently enhanced from the two paths $11\rightarrow9\rightarrow5$ and $11\rightarrow7\rightarrow5$, while the coupling between mode 1 and $1'$ is still barely affected. As we show in Fig.~\ref{fig:e4_TE}, a similar flipping of the outcoupling direction of mode 1 is observed when $\e_4$ varies from $-0.01$ to $0.01$, while those of modes $1'$ and $1''$ stay roughly the same.

\section{Conclusion}
\label{sec:conclusion}

In summary, we have shown a convenient approach to achieve and control multimode coupling using boundary wave scattering. The examples given are for solutions of the scalar Helmholtz equation with two types of open boundary conditions in a quasi-circular system, which can be realized, for example, in a liquid-jet column \cite{An}, where fine tuning of the harmonic boundary deformation has been demonstrated. The general principle should also apply for a wide variety of Hamiltonians in other geometries as well, unless the scattering is prohibited by topological property of the material \cite{edgeState1, edgeState2,edgeState3}. The boundary wave scattering presented is a linear and elastic analogue of Brillouin scattering \cite{Shen} in a circular geometry \cite{Carmon}, with the angular momentum plays the roles of frequency.
The boundary wave scattering can also couple modes within the cavity plane to propagating modes in the free space \cite{Cai}.
The cancelation of the scattering from $m=11$ to $p=5$ by the destructive interference of two scattering paths shown in Fig.~\ref{fig:e6} closely resembles the vanishing of absorption in electromagnetically induced transparency (EIT) \cite{EIT}. We also note that destructive interference between multiple scattering paths also leads to coherent backscattering in a disordered medium \cite{Akkermans}.

\vspace{0.8cm}
\section*{ACKNOWLEDGEMENTS}
We acknowledge Douglas Stone, Eugene Bogolmony, and Remy Dubertrand for helpful discussions. L.G. acknowledges MIRTHE NSF EEC-0540832. Q.S. acknowledges 2011KFB005 of the State Key Laboratory on Integrated Optoelectronics and NSFC 11204055. B.R. and H.C. acknowledge NSF ECCS-1068642 and ECCS-1128542. A.E. and J.W. acknowledge DFG research group 760.

\appendix
\section{Quasi-degeneracy and the spectral function}

\begin{figure}[t]
\centering
\includegraphics[width=0.75\linewidth]{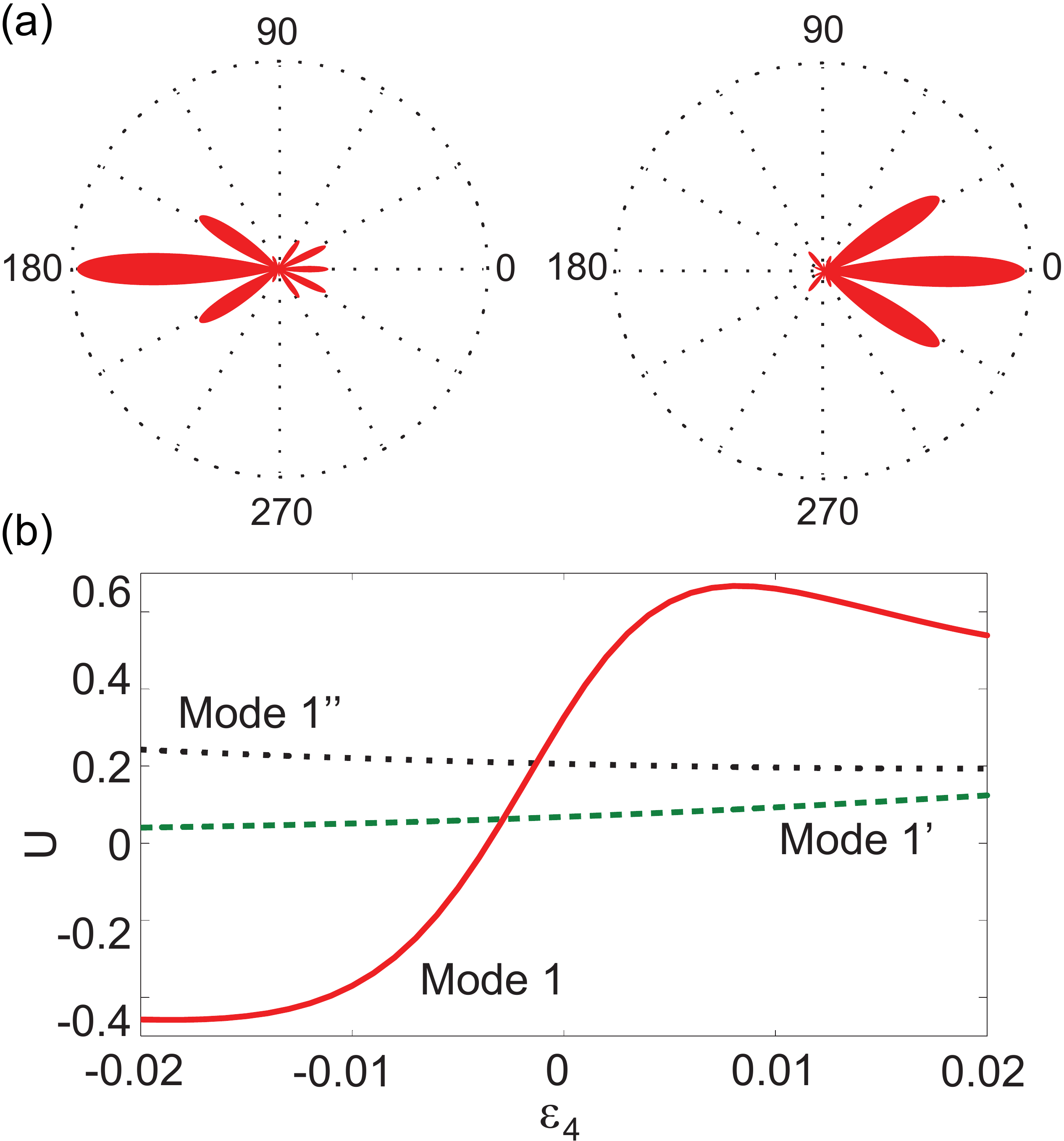}
\caption{(Color online) (a) Farfield intensity pattern of mode 1 in Fig.~\ref{fig:e6_TE} at $\e_4=-0.01$ (left) and 0.008 (right). (b) Measure of the outcoupling direction $U$ for all three modes in Fig.~\ref{fig:e6_TE} as a function of $\e_4$.
} \label{fig:e4_TE}
\end{figure}

In the appendix we discuss how the boundary wave scattering of high-$Q$ modes depends on the frequency regime. We are most interested in the $\eta=1$ modes, which have the smallest cavity decay rates and thus the lowest thresholds once optical gain is introduced ot the cavity. In Ref.~\cite{e3} it has been observed numerically that for these modes, the effect of boundary wave scattering from a minute low-order harmonic boundary deformation is most dramatic in the mesoscopic regime, i.e. the crossover regime between the wavelength regime ($\lambda\sim R$) and the semiclassical limit ($\lambda\ll R$), where $\lambda=2\pi/\re[k]$ is the wavelength; the effect becomes very weak in the semiclassical limit. Below we point out that the key to understand this phenomenon lies in the spectral function $S_p^{-1}(K_{m,1}R)$ for TM waves [or $T_p^{-1}(K_{m,1}R)$ for TE waves; see Eq.~(\ref{eq:ap_TE})], which in turn depends on th frequency spacing between $K_{m,1}$ and the nearest resonance of angular momentum $p$, as mentioned in Sec.~\ref{sec:pert}. We denote this distance $\Delta_{m,p}$, i.e.
\be
\Delta_{m,p} = \text{min}|K_{m,1}-K_{p,\eta}|\; \forall\; \eta ,
\ee
and note that it is determined mostly by the real part of the high-$Q$ resonant frequencies we are interested in, whose $|\im[KR]|\ll\re[KR]$. Thus $\re[\Delta_{m,p}]\approx\Delta_{m,p}$ is the quantity we will focus on here.

\begin{figure}[t]
\centering
\includegraphics[width=\linewidth]{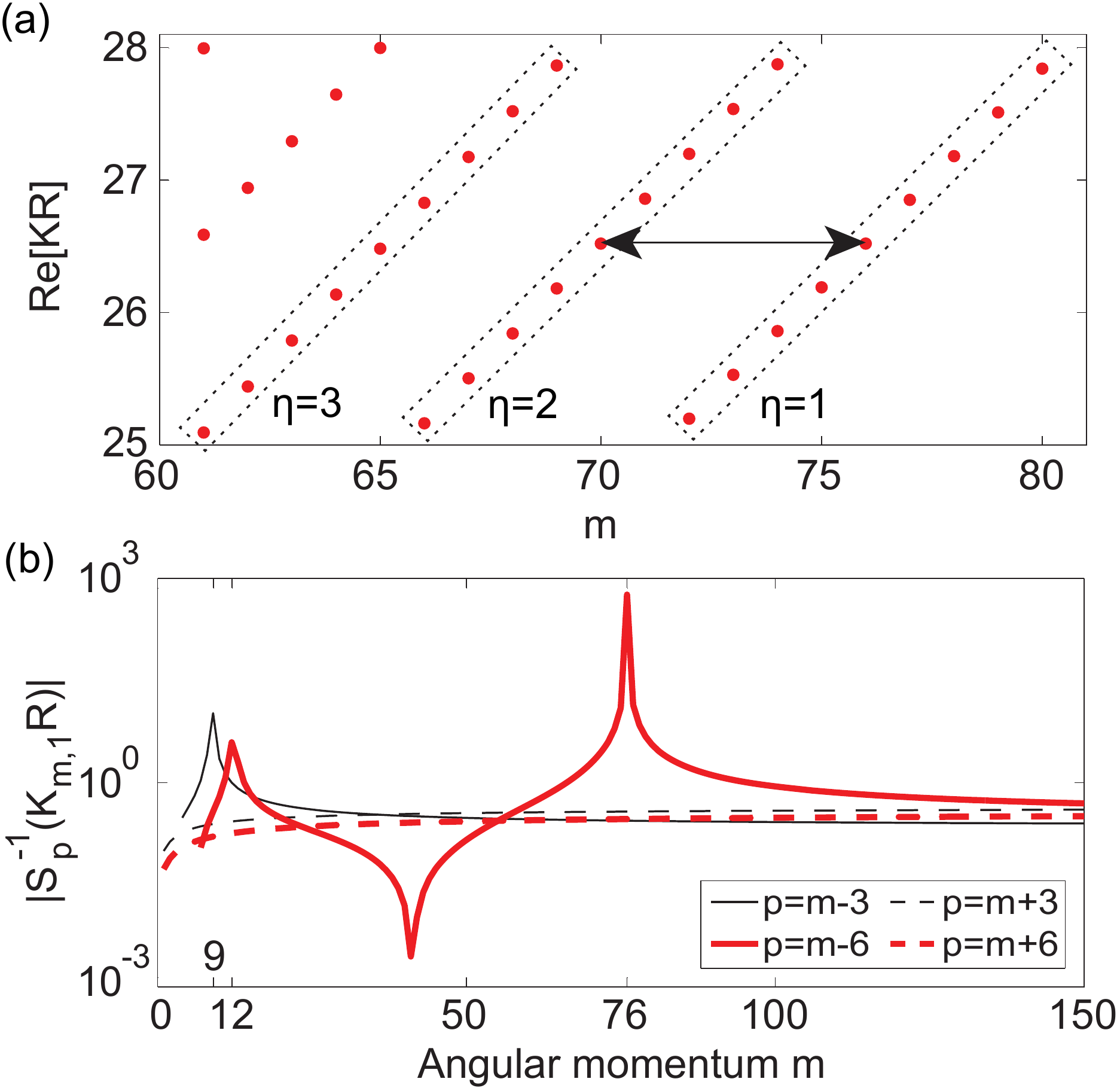}
\caption{(Color online) (a) TM spectrum of a circular cavity with refractive index $n=3.13$ near $\re[KR]=27$. Dashed-line boxes indicate the first three bands with radial quantum number $\eta=1,2,3$. $K_{m=76,\eta=1}$ and $K_{m=70,\eta=2}$ are marked by the horizontal arrows. They are the closest quasi-degenerate pair in the frequency range. (b) Spectral function $|S^{-1}_{m\pm \nu}(K_{m,1}R)|$ for the first-order scattering $m\rightarrow m\pm \nu$ as a function of $m$ for $\nu=3$ and 6.
The peak of $S^{-1}_{m-6}(K_{m,1}R)$ at $m=76$ is due to the quasi-degeneracy shown in (a). Its other peak at $m=12$ is due to another pair of quasi-degenerate modes $K_{12,1}$ and $K_{6,3}$. The single peak of $S^{-1}_{m-3}(K_{m,1}R)$ at $m=9$ is due the quasi-degeneracy between $K_{9,1}$ and $K_{6,2}$.
} \label{fig:S}
\end{figure}

To find the frequency dependence of $\Delta_{m,p}$, we first note that all $K's$ of the same $\eta$ form a band in the $\{m,\re[KR]\}$ plane [see Fig.~\ref{fig:S}(a)]. These $\eta$-bands do not cross and the slope of the $\eta=1$ band is well approximated by $\re[KR]/m\approx1/n$ \cite{JansThesis}. It is straightforward to find that $\Delta_{m,p}$ is capped at about $|m-p|/n$ for a given $m$ and $p$, which is the distance between $K_{m,1}$ and $K_{p,1}$. For $p>m$, this is in fact the value of $\Delta_{m,p}$, since $K_{p,\eta>1}$ are further away from $K_{m,1}$ as can be seen in Fig.~\ref{fig:S}(a). For such a relatively large $\Delta_{m,p}$, the spectral function $S_p^{-1}(K_{m,\eta}R)$ is typically subunitary [see Fig.~\ref{fig:S}(b)] and the associated scattering processes, such as the first-order scattering $m\rightarrow p$, are very weak. For $p<m$ however, $K_{p,\eta>1}$ can be much closer to $K_{m,1}$ when compared with $K_{p,1}$, which then leads to a very large spectral function and a very strong scattering strength. To find out when this situation occurs, we employ the approximation for $\re[KR]$ given in Ref.~\cite{Bogomolny}, which applies to both TM and TE modes of a small $\eta$:
\begin{align}
\re[KR] &= \frac{m}{n} + \frac{\beta_\eta}{n}\left(\frac{m}{2}\right)^{1/3} - \frac{1}{\tau\sqrt{n^2-1}} + O\left(\frac{1}{m}\right)^{1/3}.
\end{align}
Here $\tau=1$ for TM modes and $n^2$ for TE modes, and $\beta_\eta$ is the $\eta$-th zero of the Airy function, the first three of which are 2.34, 4.09, and 5.52. $\Delta_{m,p}$ can then be approximated by
\begin{align}
\Delta_{m,p} &\approx \text{min}\Bigg| \frac{m-p}{n} + \frac{1}{n}\left[\beta_1\left(\frac{m}{2}\right)^{1/3} - \beta_\eta\left(\frac{p}{2}\right)^{1/3}\right]\Bigg|, \nonumber \\
&= \text{min}\Bigg|\frac{m-p}{n} + \frac{1}{n}\bigg[\frac{\beta_\eta 2^{-1/3}(m-p)}{m^{2/3} + m^{1/3}p^{1/3} + p^{2/3}} \nonumber \\
 &\quad - (\beta_\eta-\beta_1)\left(\frac{m}{2}\right)^{1/3}\bigg]\Bigg|, \nonumber \\
&\approx \frac{1}{n}\text{min}\left|(m-p) - (\beta_\eta-\beta_1)\left(\frac{m}{2}\right)^{1/3}\right|.
\end{align}
for $m-p\ll m^{2/3}$. Therefore, we see that whenever
\be
L \equiv \frac{m-p}{\left(\frac{m}{2}\right)^{1/3}}+\beta_1
\ee
approaches one $\beta_{\eta>1}$, $\Delta_{m,p}$ approaches zero and the spectral function enhances the scattering strength. From this expression, we can then estimate the upper bound of $m$ for this to occur at a given $\nu\equiv m-p>0$, i.e.
\be
m_\text{max} \approx 2\left[\frac{\nu}{\beta_2-\beta_1}\right]^3 \approx 0.37\nu^3,\label{eq:quasiDeg}
\ee
which is independent of the polarization and the refractive index. For example, $m_\text{max}$ from Eq.~(\ref{eq:quasiDeg}) is $10$ and $81$ for $\nu=3,6$, respectively, which agrees qualitatively with the numerical value of $12$ and $76$ shown in Fig.~\ref{fig:S}(b). For $m>m_\text{max}$, the spectral function $S_{m-\nu}^{-1}(K_{m,1}R)$ tails off and eventually becomes comparable to the small $S_{m+\nu}^{-1}(K_{m,1}R)$ we have discussed, since now $K_{m-\nu,1}$ is the closest resonance of angular momentum $m-\nu$ to $K_{m,1}$ and $\Delta_{m,m-\nu}\approx\nu/n$, just as $K_{m+\nu,1}$ is the closest resonance of angular momentum $m+\nu$ to $K_{m,1}$ and $\Delta_{m,m+\nu}\approx\nu/n$.

Eq.~(\ref{eq:quasiDeg}) explains why the sensitivity of the high-$Q$ modes of $\eta=1$ on a low-order harmonic boundary deformation maximizes in the mesoscopic regime and becomes weak in the semiclassical limit. We note that quasi-degeneracy also occurs among modes of much larger $\eta's$, such as the TM resonances $K_{50,10}=30.187-9.6157\times10^{-6}i$ and $K_{47,11}=30.186-9.6122\times10^{-6}i$ in a circular cavity of $n=3.13$. We do not study them here because their relatively low quality factors make them difficult to observe experimentally. Their coupling, nevertheless, gives an alternative explanation to the contrasting intracavity and farfield intensity patterns found in Ref.~\cite{Creagh}, similar to what we have shown in Figs. \ref{fig:e6_outside} and \ref{fig:U_e6_TE}.



\begin{thebibliography}{99}
\bibitem{bibnotes1} In our notation the angular momentum is non-negative, and the locations of the sidebands should be understood as $|m\pm \nu|$, $|m\pm 2\nu|$, $\ldots$ We have dropped the absolute value symbols for convenience.
\end{thebibliography}
\end{document}